\documentclass[a4paper,onecolumn,11pt,unpublished]{quantumarticle}
\pdfoutput=1
\usepackage[english]{babel}
\usepackage{amsmath,amssymb,mathtools,graphicx,booktabs,microtype,placeins}
\usepackage[numbers,sort&compress]{natbib}
\usepackage[hidelinks]{hyperref}

\graphicspath{{figures/}}
\newcommand{\ii}{\mathrm{i}}
\newcommand{\dd}{\mathrm{d}}
\newcommand{\CP}{\mathbb{CP}}
\newcommand{\norm}[1]{\lVert #1\rVert}
\newcommand{\fig}[4]{\begin{figure}[!htb]\centering\includegraphics[width=#2\linewidth]{#1}\caption{#3}\label{#4}\end{figure}}

\begin{document}

\title{Probing an exceptional-point Jordan chain with heralded photonic textures}
\author{Sheng-Hong Lai}
\email{shlai@cycu.edu.tw}
\thanks{These authors contributed equally to this work.}
\author{Wen-Yu Wen}
\email{wenw@cycu.edu.tw}
\thanks{Corresponding author. These authors contributed equally to this work.}
\affiliation{Department of Physics and Center for High Energy Physics, Chung Yuan Christian University, Taiwan}
\maketitle

\begin{abstract}
We propose a two-photon protocol that probes the Jordan chain of a third-order exceptional point (EP3) through the order of an optical dark point. Measuring one photon at transverse position \(w\) prepares its partner with relative mode amplitudes \([1:w:w^2]\). Integrating the Berry curvature of this texture in units of \(2\pi\) gives charge two. At fixed propagation time, interference with an identity reference gives a filter with an amplitude zero of order one after one stage and order two after two stages. The conditional states approach well-defined rays at the dark point, and their extended textures have charges one and zero, respectively. The lost charge equals the zero multiplicity, or equivalently the length of the local sheaf quotient. A small cyclic perturbation \(\delta\) lifts the zero and restores total charge two. The recovered charge concentrates within a radius of order \(|\delta|\) after one stage and \(|\delta|^{1/2}\) after two stages. Synthetic coincidence counts illustrate the measurement cost of resolving these shrinking regions.
\end{abstract}

\section{Introduction}
A position measurement on one photon of an entangled pair can prepare a spatial pattern of quantum states in the other photon. In a nonlocal optical skyrmion, the first photon's position labels the pattern and the second photon's mode state supplies its geometry. Coincidence measurements give access to the resulting texture \cite{r1}. Optical skyrmions have also been studied in local fields \cite{r2,r3,r4,r5,r6,r7} and as probes of noise and quantum channels \cite{r8,r9}. Here we use a three-mode photonic texture to probe the successive steps of a non-Hermitian Jordan chain.

At a third-order exceptional point (EP3), three eigenvalues and their eigenvectors coalesce. The resulting length-three Jordan chain contains one eigenvector and two generalized eigenvectors \cite{r10,r11,r12,r13}. The finite-time propagator is invertible even at the EP3, so applying it uniformly to the texture preserves the total charge. Our model places the coalesced eigenvalue at zero. Interfering the propagated amplitude with an identity reference exposes the Jordan chain. At the difference port, the reference cancels the amplitude of the EP eigenvector. Cascading this port cancels one additional input direction at each stage. A position-dependent input then translates the sequence of cancelled directions into a sequence of spatial amplitude zeros.

Eigenframe textures, state transport around closed parameter loops, and spectral winding measurements probe distinct aspects of non-Hermitian topology \cite{r14,r15,r16,r17}. The geometric quantity here is the Berry curvature of the conditional photon state as a function of transverse position. At each generator setting, the position measurement scans that texture with the generator and propagation time held fixed.

\subsection{Main results and scope}
For the qutrit texture \([1:w:w^2]\), matched reference subtraction exposes the nilpotent part of the EP3 generator \cite{r18}. The first two stages give amplitude zeros of orders one and two. The missing charge is exactly that order, so the total charge of the extended texture falls from two to one and then to zero. The same integer is the length of a local quotient that measures the vanishing of the optical map at the dark point. Small cyclic detuning returns this charge to the texture and concentrates it near the center. The radius and the probability of successful transmission together determine the counting cost of resolving the concentration.

We assume a pure conditional input, coherent cascaded stages, and a fixed nonzero propagation time. Photon A heralds the input state by its position, and a coincidence at the selected port of photon B heralds successful filtering. A coherent classical field with the same mode amplitudes has the same ray geometry. Identifying the measured pattern with a particular Jordan chain relies on independent calibration of the signal arm and the reference. We give a conceptual optical implementation and use synthetic counts to illustrate its measurement limits.

\section{Conditional texture and optical readout}\label{sec:setup}
\subsection{State preparation and Berry charge}
The position measurement on photon A supplies the dimensionless coordinate \(w=x+\ii y\). Photon B occupies three orthonormal modes \(e_1,e_2,e_3\). We write the biphoton amplitude as \(\chi(w)f(w)\), where \(f(w)\) is the column of conditional mode amplitudes and \(S_{\rm in}(w)\) is its squared norm.
\begin{equation}
 f(w)=(1,w,w^2)^{\mathsf T}, \qquad S_{\rm in}(w)=1+|w|^2+|w|^4.
 \label{eq:input}
\end{equation}
The envelope \(\chi\) controls the position count rate. A smooth, nonzero, rapidly decreasing envelope such as a Gaussian makes the pair state normalizable. After normalization and removal of the irrelevant overall phase, each position selects a ray \([f(w)]\) in \(\CP^2\), the three-mode analogue of the two-mode Poincar\'e sphere.

As the position varies, the relative amplitudes and phases of the conditional state change. For a nonvanishing amplitude column \(F\), write \(u=F/\norm{F}\) and use the Berry connection \(\mathcal A=-\ii u^\dagger\dd u\). The curvature \(\dd\mathcal A\) is invariant under a position-dependent phase of \(u\). Its flux through a small disc gives the geometric phase around the boundary. We call the curvature integral divided by \(2\pi\) the charge. We use the orientation \(\dd x\wedge\dd y\). The charge over a finite disc \(|w|<R\) can vary continuously with \(R\). The input ray tends to \([e_3]\) in every direction at large radius, so adding the point at infinity gives a compact sphere and an integer total charge of two. Figure~\ref{fig:input}(b) shows how the finite-disc charge approaches this integer. Appendix~\ref{app:geometry} gives the boundary formula used to compute it.

\subsection{Heralded filtering and coincidence readout}
In the signal arm of Fig.~\ref{fig:setup}, propagation for a fixed interval \(\tau\ne0\) implements \(U_\delta\). The generator, propagator, and \(k\)-stage transfer at the selected port are
\begin{equation}
 J_\delta=\begin{pmatrix}0&1&0\\0&0&1\\\delta&0&0\end{pmatrix},
 \qquad U_\delta=e^{-\ii\tau J_\delta},\qquad
 T_{k,\delta}=s_0^k(U_\delta-I)^k .
 \label{eq:transfer}
\end{equation}
The unit nearest-neighbor coupling sets the time scale. The perturbation \(\delta\) couples \(e_1\) back to \(e_3\), closing the cycle. At each balanced interferometer, the reference arm implements the identity on the three modes and the selected output subtracts its amplitude from that of the signal arm. The calibrated field factor \(s_0>0\) sets the overall transmission. Success is declared at the selected output of the last stage, with the intermediate stages connected coherently.

\fig{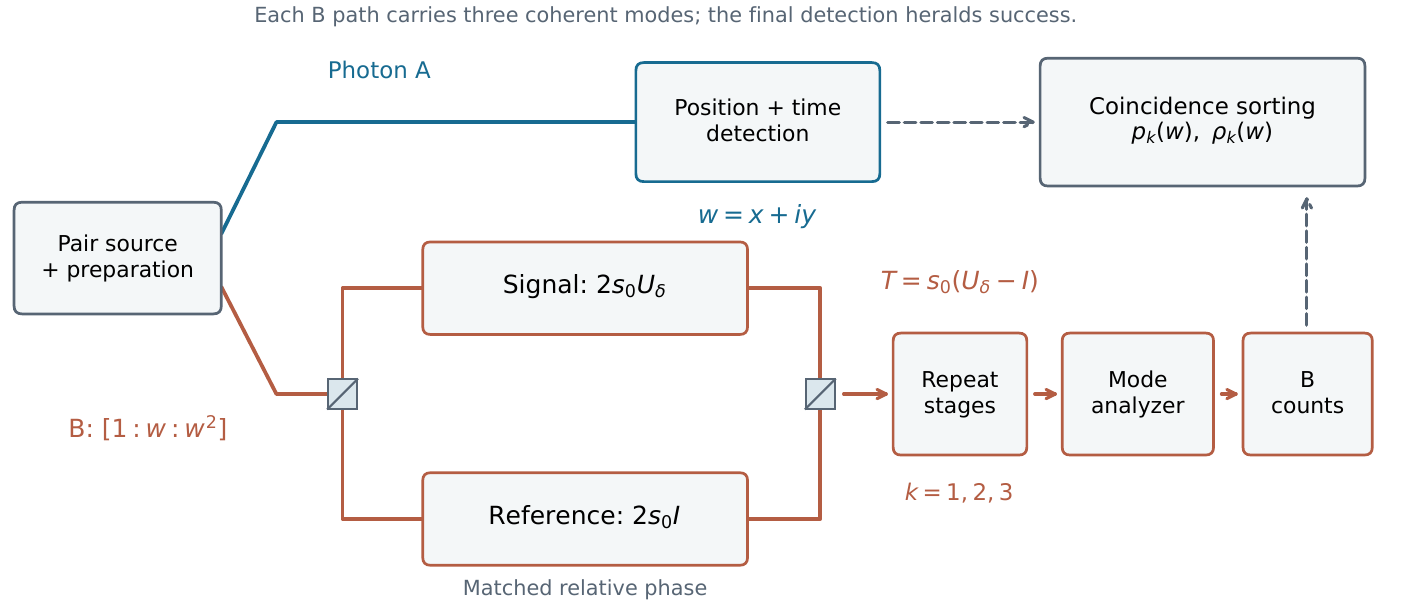}{1}{Proposed two-photon measurement. Detection of photon A gives \(w\) and the timing signal for coincidence sorting. Photon B passes through the selected difference ports of \(k\) identical, coherently connected interferometers and then through a mode analyzer. The coincidence rate relative to the incident pair rate estimates the success probability. An informationally complete set of analyzer outcomes determines the conditional output state.}{fig:setup}

In quantum-information terms, \(T_{k,\delta}\) is a Kraus operator for the successful outcome of a completely positive, trace-nonincreasing map. The unnormalized output for an input density operator \(\rho\) is \(T_{k,\delta}\rho T_{k,\delta}^\dagger\). For the pure input in Eq.~\eqref{eq:input}, define \(F_{k,\delta}=T_{k,\delta}f\). Then \(p_k(w)\) is the success probability and \(\rho_k(w)\) is the state conditioned on success. The latter is defined where \(p_k(w)>0\).
\begin{equation}
 p_k(w)=\frac{\norm{F_{k,\delta}(w)}^2}{S_{\rm in}(w)},
 \qquad
 \rho_k(w)=\frac{F_{k,\delta}(w)F_{k,\delta}(w)^\dagger}{\norm{F_{k,\delta}(w)}^2}.
 \label{eq:conditional}
\end{equation}
Efficiency-corrected coincidences summed over analyzer outcomes estimate \(p_k\), while their relative frequencies among successful events determine \(\rho_k\). This distinction is essential at a dark point, where transmission can vanish even when the output ray approaches a unique limit. Appendix~\ref{app:filter} derives the passive implementation bound and the coincidence probabilities.

\section{Dark points at the tuned EP3}\label{sec:tuned}
At \(\delta=0\), \(J_0\) sends \(e_3\) to \(e_2\), \(e_2\) to \(e_1\), and \(e_1\) to zero. Propagation leaves the \(e_1\) eigenvector unchanged, so the matched reference cancels it at the difference port. With \(\zeta=-\ii\tau\), the exact transmitted amplitudes are
\begin{equation}
 \begin{aligned}
 F_{1,0}&=s_0\zeta w(1+\zeta w/2,w,0)^{\mathsf T},\\
 F_{2,0}&=s_0^2\zeta^2 w^2 e_1,\qquad F_{3,0}=0 .
 \end{aligned}
 \label{eq:tuned}
\end{equation}
The input at the center is \(e_1\), so the first two stages both give a dark point at \(w=0\). Just off center, the \(we_2\) input can feed \(e_1\) at the first difference port. A second stage cancels that contribution, leaving only the \(w^2e_3\) input that can undergo two successive mode conversions. The coincidence probability therefore rises as \(|w|^2\) after one stage and as \(|w|^4\) after two. Figure~\ref{fig:input}(c) shows both slopes. A third stage extinguishes the selected output everywhere.

Away from the center, the scalar factors \(w\) and \(w^2\) in Eq.~\eqref{eq:tuned} leave the normalized output ray unchanged. Dividing them out gives a smooth extension through the dark point, with limiting ray \([e_1]\) in both cases. After one stage the extended ray has a linear polynomial representative and carries charge one. After two stages it is constant and carries charge zero. The extension is inferred from nearby positions because the success probability at the point itself is zero. Table~\ref{tab:tuned} collects the tuned results.

\begin{table}[!htbp]
\centering
\begin{tabular}{cccc}
\toprule
Stages & Common factor & Near-center success probability & Total charge\\
\midrule
0 & \(1\) & \(1\) & 2\\
1 & \(w\) & \(\propto |w|^2\) & 1\\
2 & \(w^2\) & \(\propto |w|^4\) & 0\\
3 & zero field & zero everywhere & undefined\\
\bottomrule
\end{tabular}
\caption{Tuned EP3 with the aligned input and \(\tau\ne0\). The total charge is that of the input at stage zero and of the extended output ray at stages one and two.}
\label{tab:tuned}
\end{table}

\fig{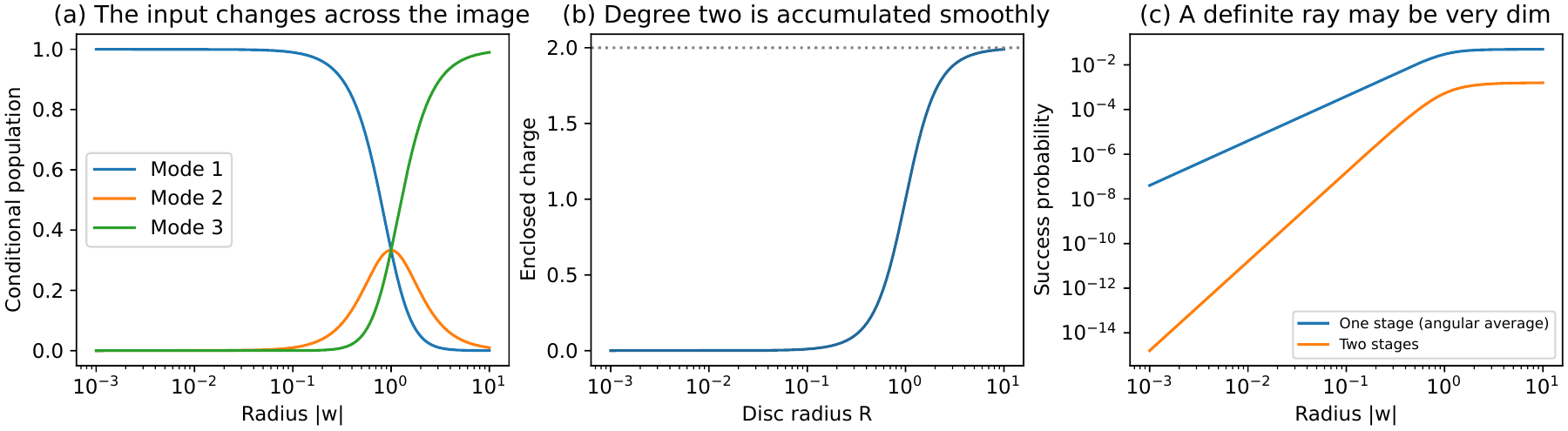}{1}{Input texture and tuned transmission. (a) Populations of the three modes in the normalized input \([1:w:w^2]\). (b) Berry charge of the input inside a disc of radius \(R\). (c) Success probabilities after one and two stages at \(\tau=1\) and \(s_0=0.2\). The one-stage curve is averaged over the polar angle.}{fig:input}

On a small circle around the dark point, the common amplitude factor winds in phase once after one stage and twice after two. The winding equals the decrease in integrated Berry curvature, in units of \(2\pi\), between the input and the extended output. Appendix~\ref{app:geometry} derives the relation using the Berry connection. Appendix~\ref{app:sheaf} shows why the same integer is the length of the sheaf quotient supported at the dark point. This length counts the local Taylor coefficients excluded by the vanishing optical map.

\section{Charge concentration under cyclic detuning}\label{sec:detune}
At fixed \(\tau\), a sufficiently small nonzero \(\delta\) makes the selected transfer invertible. The output has a quadratic polynomial representative with no common zero and therefore again has total charge two. To locate the restored charge, consider the neighborhood of \(w=0\). The perturbation contributes an amplitude proportional to \(\delta\) in a mode combination independent of the leading tuned amplitude. That tuned amplitude is proportional to \(w\) after one stage and to \(w^2\) after two. Balancing their magnitudes gives core radii of order \(|\delta|\) and \(|\delta|^{1/2}\). The eigenvalues of \(J_\delta\) split on the distinct scale \(|\delta|^{1/3}\). The spatial exponents come from the competition between the two terms in the transmitted state (Appendix~\ref{app:core}).

Inside the core the perturbation dominates and the conditional ray is nearly independent of angle. Outside the core the tuned amplitude dominates. Where the amplitudes are comparable, the mode composition changes with radius and the relative phase winds \(k\) times on a circle for \(k=1,2\). The combined radial and angular variation produces the Berry curvature associated with the recovered charge.

Let \(Q_k(R,\delta)\) be the charge inside a centered disc for the aligned input. The one-stage core carries one unit, with the other unit in the outer texture. After two stages the core carries both units. Define the half-charge radius by \(Q_k(R_{1/2}^{(k)},\delta)=k/2\), half of the asymptotic core contribution. At \(\tau=1\),
\begin{equation}
 R_{1/2}^{(1)}\sim\sqrt{23/18}\,|\delta|,
 \qquad
 R_{1/2}^{(2)}\sim 2^{1/4}|\delta|^{1/2}.
 \label{eq:radii}
\end{equation}
In Fig.~\ref{fig:charge}(a,b), decreasing \(|\delta|\) by a factor of 100 reduces the half-charge radii by factors of about 100 and 10, respectively. Figure~\ref{fig:collapse} shows how the curves approach a common profile for each stage after rescaling the radius. Figure~\ref{fig:radii} compares the exact half-charge radii with the two asymptotic power laws.

\fig{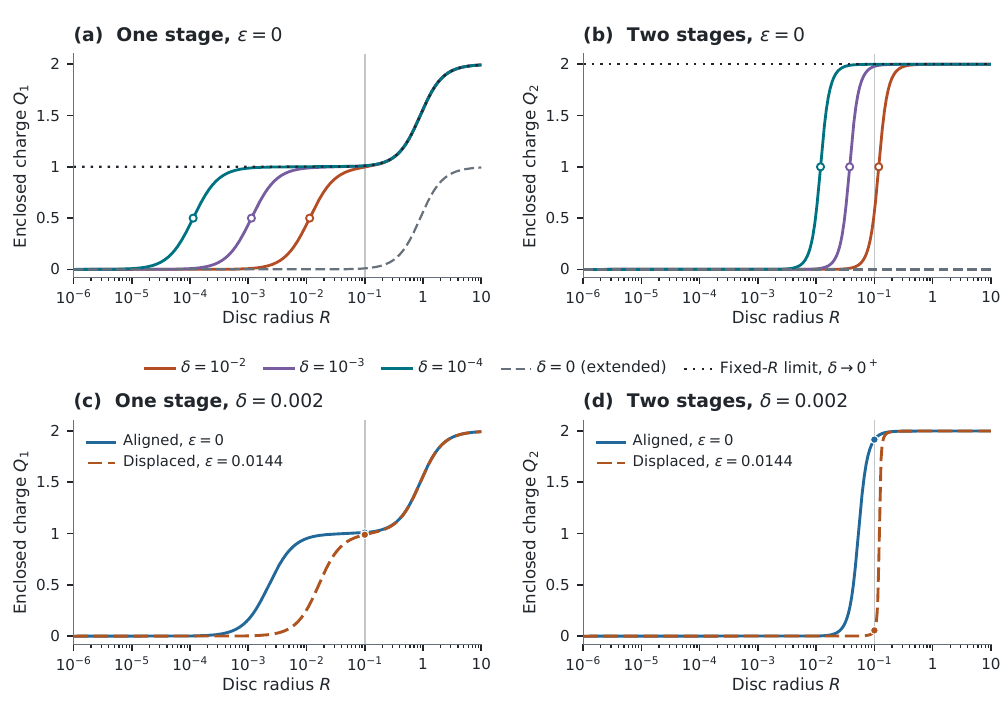}{1}{Charge within a centered disc at \(\tau=1\). The left and right columns show one and two stages. (a,b) Aligned input at three positive detunings. Open circles mark half the core charge. Dashed curves give the exactly tuned extended charge, and dotted curves give the limit at fixed \(R\) through positive detunings. (c,d) Aligned and displaced inputs at \(\delta=0.002\). Filled points give the charges at \(R=0.1\), the aperture used in Fig.~\ref{fig:counts}.}{fig:charge}

For any fixed \(R>0\), taking \(\delta\) to zero through nonzero values gives the charge of the tuned extended ray in that disc plus \(k\). In this limit the additional curvature becomes concentrated at the origin. At exact tuning the extended ray is smooth there. These two cases give the distinct curves shown in Fig.~\ref{fig:charge}.

The role of the input can be seen by preparing \([1:w^m:w^n]\) for integers \(0<m<n\). Its input charge is \(n\). The aligned tuned filters give common factors \(w^m\) and \(w^n\), extended charges \(n-m\) and zero, and near-center success probabilities proportional to \(|w|^{2m}\) and \(|w|^{2n}\). Small nonzero cyclic detuning restores total charge \(n\) in each case. The cores then carry \(m\) and \(n\) units with respective radii of order \(|\delta|^{1/m}\) and \(|\delta|^{1/n}\).

\section{Alignment, calibration, and counting limits}\label{sec:controls}
\paragraph{Preparation alignment.}
A constant offset in the third mode gives \(f_\epsilon=(1,w,w^2+\epsilon)^{\mathsf T}\). At the tuned EP3, a one-stage dark point requires \(w=0\) and \(w^2+\epsilon=0\) at the same position. An offset \(\epsilon\ne0\) removes the common zero and returns the total charge to two. After two stages the common factor is \(w^2+\epsilon\). The offset splits the double zero into two simple zeros at \(w=\pm\sqrt{-\epsilon}\) while the extended ray remains constant. Near each split zero the success probability is locally quadratic.

Figure~\ref{fig:charge}(c,d) compares the aligned and displaced inputs at \(\delta=0.002\). For one stage, the \(R=0.1\) disc encloses close to one unit in both cases. For two stages it encloses about \(1.916\) units for the aligned input and \(0.057\) for \(\epsilon=0.0144\). In the displaced case the two zeros at exact tuning lie at \(\pm0.12\ii\), outside that aperture. The comparison shows how alignment affects the charge inferred from a finite field of view.

\paragraph{Reference calibration and interpretation.}
A complex amplitude mismatch \(\eta\) replaces the reference \(I\) with \((1+\eta)I\). Even with the generator tuned to the EP3, a nonzero \(\eta\) makes the difference-port transfer invertible and produces success probability \(s_0^{2k}|\eta|^{2k}\) at the original dark point. Independent amplitude and phase calibration is therefore required to attribute a lifted zero to cyclic detuning. A static optical filter synthesized from the singular-value decomposition of \(T_{k,0}\) can reproduce the same success probabilities and conditional states. The readout characterizes the selected transfer, and independent calibration ties that transfer to the EP3 generator.

\paragraph{Counting limits.}
Within a core-sized aperture, the mean success probability scales as \(|\delta|^2\). For a smooth incident beam of fixed transverse extent, the fraction of input pairs entering that aperture scales as its area, of order \(|\delta|^{2/k}\). Thus the signal yield per incident pair scales as \(|\delta|^{2+2/k}\). Charge can remain finite while the expected number of usable coincidences falls sharply.

In the synthetic two-stage example at aligned \(\delta=0.002\), two million trials distributed over a position grid give 26 selected coincidences, including background. Fitting those counts to the calibrated transfer model gives a charge interval spanning essentially zero to two for the \(R=0.1\) disc. Figure~\ref{fig:counts} compares such intervals with the exact model curves. This is a model-based inference from success counts. Reconstructing the state independently would use mode-resolved coincidences as described in Section~\ref{sec:setup}. Appendix~\ref{app:counts} specifies the exposure and inference procedure.

\fig{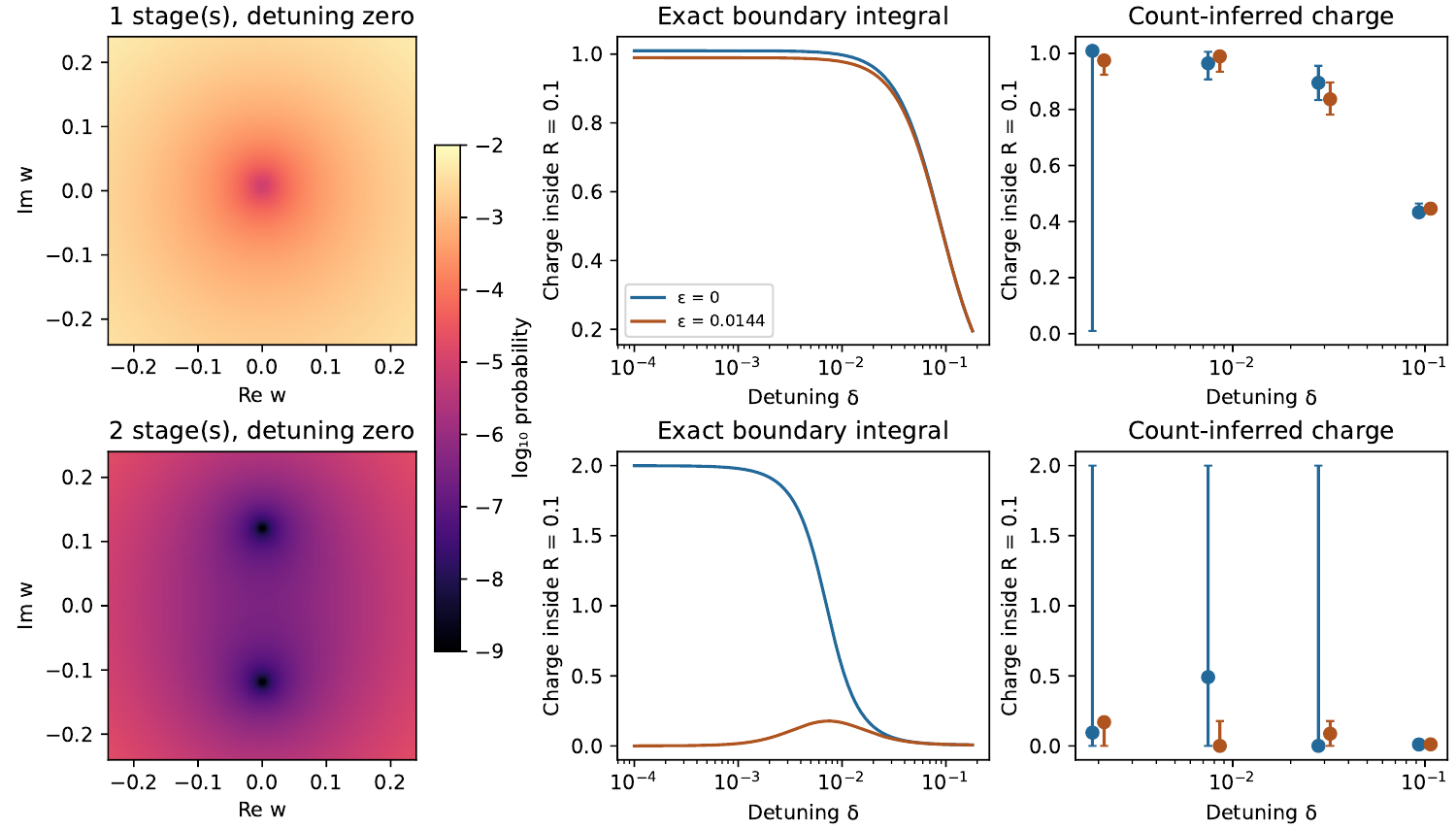}{1}{Synthetic count analysis at \(\tau=1\) and \(s_0=0.2\). Rows show one and two stages. The left panels show tuned success-probability maps for \(\epsilon=0.0144\) on the same logarithmic scale. The middle panels give the exact model charge inside \(R=0.1\) for aligned and displaced inputs. The right panels show charge estimates from synthetic coincidences using the calibrated transfer model. Bars span the minimum and maximum charges among detunings in each 95\% Monte Carlo confidence set on the fitting grid. Circles and squares indicate the aligned and displaced inputs, respectively, with small horizontal offsets for visibility.}{fig:counts}

\section{Conclusion}
Matched reference subtraction converts the Jordan hierarchy of a calibrated EP3 into an order of vanishing that can be read from coincidence probabilities. For the input \([1:w:w^2]\), a simple and a double dark point account for the loss of one and two units of Berry charge, respectively. The local quotient length gives the same integer as the geometric winding. Cyclic detuning restores the total charge and places the recovered curvature on distinct spatial scales after the two stages. The finite-aperture and counting analysis connects these predictions to the spatial resolution and coincidence rates the optical protocol would require.

\paragraph{Acknowledgements.}
WYW is supported in part by the National Science and Technology Council of Taiwan under Grants No.~115-2112-M-033-006 and 115-2112-M-033-010. SHL is supported in part under Grant No.~114-2112-M-033-014-MY3.

\paragraph{Data and code availability.}
The source archive includes the script for Fig.~\ref{fig:counts}, its random seed, the simulated bin counts, and the fitting grid with charge values and Monte Carlo test results. The counts are synthetic.

\paragraph{Author contributions and AI disclosure.}
Sheng-Hong Lai and Wen-Yu Wen contributed equally to this work. ChatGPT assisted with manuscript organization, drafting and language editing, numerical checks, and preparation of the synthetic-count simulation and figure code. The authors take responsibility for the final manuscript.

\FloatBarrier
\appendix
\section{Projective curvature and the index of a common zero}\label{app:geometry}
\paragraph{Charge from the mode amplitudes.}
Let \(F\) be a holomorphic amplitude column that is nonzero in the region of interest. Write \(S=F^\dagger F\), \(F'=\partial_w F\), and \(u=F/\sqrt S\). Then
\[
 u^\dagger\dd u=\frac{F^\dagger\dd F}{S}-\frac{\dd S}{2S},
 \qquad \partial_w S=F^\dagger F',
 \qquad \partial_{\bar w}\partial_w S=\norm{F'}^2 .
\]
With \(\Delta=4\partial_w\partial_{\bar w}\), the connection \(\mathcal A=-\ii u^\dagger\dd u\) gives
\begin{equation}
 \frac{\dd\mathcal A}{2\pi}
 =\frac{\Delta\log S}{4\pi}\dd x\wedge\dd y
 =\frac{S\norm{F'}^2-|F^\dagger F'|^2}{\pi S^2}\dd x\wedge\dd y.
 \label{eq:curv}
\end{equation}
The last expression is nonnegative by the Cauchy--Schwarz inequality. If \(F\) is regular and nonvanishing on the disc and its boundary, Stokes' theorem gives
\begin{equation}
 Q_F(R)=\frac1{2\pi}\int_0^{2\pi}
 \frac{\operatorname{Re}(F^\dagger wF')}{\norm{F}^2}\,\dd\theta,
 \qquad w=Re^{\ii\theta}.
 \label{eq:boundary}
\end{equation}
For the input, this gives \(Q_f(R)=(R^2+2R^4)/(1+R^2+R^4)\). More generally, a degree-\(d\) polynomial column without common finite zeros has total charge \(d\). Its boundary integrand tends uniformly to \(d\) as \(R\to\infty\), and \(v^dF(1/v)\) is a nonzero local representative at the point \(v=1/w=0\). For a column with a common zero, apply Eq.~\eqref{eq:boundary} to the extended ray after cancelling the common factor.

\paragraph{Contribution of an isolated dark point.}
Let \(K\) be the selected transfer. Near an isolated zero, take smooth unit vectors \(u_{\rm in}\) and \(\tilde u\) for the input and extended output rays and write \(Ku_{\rm in}=g\tilde u\). On the punctured disc, the normalized transmitted vector is \(e^{\ii\arg g}\tilde u\). Its Berry connection differs from that of \(\tilde u\) by \(\dd\arg g\). Away from the zeros, the transmitted and input vectors transform by the same phase under a change of the input representative, so the difference of their connections is globally defined. Integrate its exterior derivative over the compact surface with a small disc removed around each zero. The holes are clockwise boundaries of the remaining surface. The integrals of the smooth connection terms vanish as the discs shrink, leaving
\begin{equation}
 Q_{\rm in}-Q_{\rm ext}=\sum_p \frac1{2\pi}\oint_{\partial D_p}\dd\arg g ,
 \label{eq:index}
\end{equation}
where the small circles \(\partial D_p\) are oriented counterclockwise. This winding is the order of a holomorphic zero. For the aligned EP3, the output is regular and nonzero in the chart at infinity after the appropriate rescaling of its polynomial representative. The entire deficit therefore comes from the dark point at \(w=0\).

\section{Selected transfer and conditional measurement}\label{app:filter}
\paragraph{Passive implementation and detected events.}
A balanced splitter contributes a factor \(1/\sqrt2\) at the entrance to each arm and a second factor at recombination. With arm transfer matrices \(2s_0U_\delta\) and \(2s_0I\), the difference port therefore implements \(s_0(U_\delta-I)\). Both arms are passive provided
\[
 0<s_0\le [2\max\{1,\norm{U_\delta}_{\rm op}\}]^{-1}.
\]
For a mismatched reference \((1+\eta)I\), the corresponding conditions are \(2s_0|1+\eta|\le1\) and \(2s_0\norm{U_\delta}_{\rm op}\le1\). The selected transfer is thus a contraction. It gives one outcome of a quantum instrument, with the squared norm of the output giving the success probability \cite{r20}.

The pair state and its normalization are
\[
 |\Psi\rangle=\mathcal N\int \dd^2w\,\chi(w)|w\rangle_A\otimes f(w)_B,\qquad
 |\mathcal N|^{-2}=\int|\chi|^2S_{\rm in}\,\dd^2w.
\]
The incident position density is \(I_A(w)=|\mathcal N\chi(w)|^2S_{\rm in}(w)\). If an analyzer is described by positive operators \(E_j\) with \(\sum_j E_j=I\), the coincidence density before detector losses is
\[
 I_A(w)\,p_k(w)\,\operatorname{tr}\!\left[E_j\rho_k(w)\right].
\]
Summing over \(j\) gives \(I_Ap_k\). Dividing the density for each outcome by this sum yields the conditional frequencies for tomography. For the displaced input, replace \(S_{\rm in}\) by \(\norm{f_\epsilon}^2\) in both the source density and the success probability.

\paragraph{Kernels and exact amplitudes.}
Introduce the entire function
\[
 B_\tau(z)=\frac{e^{-\ii\tau z}-1}{z},\qquad B_\tau(0)=\zeta=-\ii\tau .
\]
Then \(U_\delta-I=B_\tau(J_\delta)J_\delta\). At \(\delta=0\), \(B_\tau(J_0)\) is triangular with nonzero diagonal \(\zeta\) and commutes with \(J_0\). It follows that \(\ker T_{k,0}=\ker J_0^k=\operatorname{span}\{e_1,\ldots,e_k\}\) for \(k=1,2,3\). Using \(J_0^3=0\) in the exponential gives \(U_0-I=\zeta J_0+\zeta^2J_0^2/2\) and \((U_0-I)^2=\zeta^2J_0^2\), which yields Eq.~\eqref{eq:tuned}. The tuned success probabilities are
\[
 p_1=\frac{s_0^2|\zeta|^2|w|^2}{S_{\rm in}}
 \left(|1+\zeta w/2|^2+|w|^2\right),
 \qquad p_2=\frac{s_0^4|\zeta|^4|w|^4}{S_{\rm in}}
 \quad (\delta=0).
\]
To obtain the amplitudes for the displaced input, replace \(w^2\) by \(w^2+\epsilon\) in the columns \(s_0(\zeta w+\zeta^2w^2/2,\zeta w^2,0)^{\mathsf T}\) and \(s_0^2\zeta^2w^2e_1\). A reference mismatch adds a diagonal \(-s_0\eta\) to the one-stage tuned triangular transfer, giving determinant \((-s_0\eta)^3\). For \(\delta\ne0\), \(J_\delta^3=\delta I\) and its three eigenvalues are the distinct roots of \(\lambda^3=\delta\). For \(k\ge1\) the selected transfer is invertible when \(\tau\lambda\notin2\pi\mathbb{Z}\) for every root. The sufficient condition \(0<|\tau||\delta|^{1/3}<2\pi\) gives the small-detuning regime used in the main text.

\section{Charge profiles at fixed propagation interval}\label{app:core}
\paragraph{Expansion near the dark point.}
For \(k=1,2\), the first terms of the transmitted amplitude are
\begin{equation}
 s_0^{-k}F_{k,\delta}=a_k w^k+b_k\delta+o(|w|^k+|\delta|),
 \label{eq:local}
\end{equation}
where \(B_*=B_\tau(J_0)\) and
\[
 (a_1,b_1)=(B_*e_1,B_*e_3),\qquad
 (a_2,b_2)=(B_*^2e_1,B_*^2e_2).
\]
To see this, note that \(J_\delta f=(w,w^2,\delta)^{\mathsf T}\), \(J_\delta^2f=(w^2,\delta,\delta w)^{\mathsf T}\), and \(B_\tau(J_\delta)^k=B_*^k+O(|\delta|)\) by analyticity. These give remainders of order \(O(|w|^2+|\delta w|+|\delta|^2)\) and \(O(|\delta w|+|\delta|^2)\), respectively. Each pair of leading vectors is linearly independent since \(B_*\) is invertible. At fixed \(\arg\delta\), rescale to \(w=|\delta|^{1/k}z\) and divide the column by \(|\delta|\). The limit is \(a_kz^k+b_ke^{\ii\arg\delta}\). The rescaled polynomial coefficients converge, so the column and its first spatial derivative converge uniformly on bounded discs in \(z\). The limiting column is nowhere zero, which gives convergence of the charge on each such disc. The phase of \(\delta\) shifts the angular dependence of the limit and leaves the centered disc integral invariant.

\paragraph{Limiting charge and radius.}
For the limiting column, let \(A=\norm{a_k}^2\), \(D=\norm{b_k}^2\), \(C=a_k^\dagger b_k\), and \(\sigma=R^k/|\delta|\). After the scalar rescaling, the denominator in Eq.~\eqref{eq:boundary} is \(A\sigma^2+D+2|C|\sigma\cos\psi\), where \(\psi=k\theta-\arg\delta-\arg C\). Its numerator is \(k(A\sigma^2+|C|\sigma\cos\psi)\). Subtracting \(k/2\) times the denominator from the numerator leaves an angle-independent term \(k(A\sigma^2-D)/2\). The angular mean of \((a+b\cos\psi)^{-1}\) is \((a^2-b^2)^{-1/2}\) for \(a>|b|\), so
\begin{equation}
 Q_{k,\rm core}(\sigma)=\frac{k}{2}\left[
 1+\frac{A\sigma^2-D}{\sqrt{(A\sigma^2+D)^2-4|C|^2\sigma^2}}
 \right].
 \label{eq:profile}
\end{equation}
Linear independence gives \(AD>|C|^2\), so the profile increases from zero to \(k\) and has strictly positive derivative with respect to \(v=\sigma^2\).
\[
 \frac{\dd Q_{k,\rm core}}{\dd v}
 =\frac{k(AD-|C|^2)(Av+D)}{[(Av+D)^2-4|C|^2v]^{3/2}}>0.
\]
The charge reaches \(k/2\) when \(A\sigma^2=D\). Convergence on nearby rescaled discs and monotonicity of the full charge from Eq.~\eqref{eq:curv} then give \(R_{1/2}^{(k)}\sim (D/A)^{1/(2k)}|\delta|^{1/k}\). At \(\tau=1\),
\[
 a_1=-\ii e_1,\quad b_1=\tfrac{\ii}{6}e_1-\tfrac12e_2-\ii e_3,
 \qquad a_2=-e_1,\quad b_2=\ii e_1-e_2 .
\]
The corresponding \((A,D,|C|)\) are \((1,23/18,1/6)\) and \((1,2,1)\). These give Eq.~\eqref{eq:radii} and the limiting curves in Figs.~\ref{fig:collapse} and \ref{fig:radii}.

\paragraph{Limit at fixed aperture.}
At fixed \(R>0\), the boundary column and its derivative converge to those of \(w^kH_k\), where \(H_k\) is the regular tuned column after division by \(w^k\). The boundary integrand for \(w^kH_k\) is that of \(H_k\) plus \(k\). Hence
\[
 \lim_{\delta\to0,\,\delta\ne0}Q_k(R,\delta)=k+Q_{k,\rm ext}(R).
\]
Shrinking the disc first at fixed nonzero detuning instead gives zero charge. Off the origin, including the chart at infinity, the curvature converges smoothly to that of the extended tuned ray. The excess \(k\) on every fixed centered disc gives a point mass of weight \(k\) in the weak limit of the normalized curvature. The one-stage tuned curve in Fig.~\ref{fig:charge} follows by inserting \(H_1=(1+\zeta w/2,w,0)^{\mathsf T}\) into Eq.~\eqref{eq:boundary} and averaging over the circle. The tuned two-stage curve vanishes because \(H_2\) is constant.

\renewcommand{\thefigure}{S\arabic{figure}}
\renewcommand{\theHfigure}{supp.\arabic{figure}}
\setcounter{figure}{0}
\fig{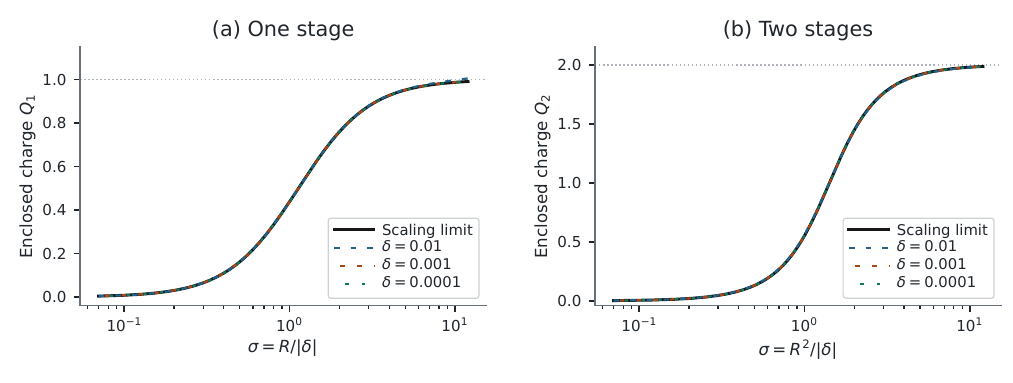}{1}{Enclosed charge as a function of the rescaled radius at \(\tau=1\). For one stage, \(\sigma=R/|\delta|\). For two stages, \(\sigma=R^2/|\delta|\). The black curves show Eq.~\eqref{eq:profile} with half-charge points at \(\sigma=\sqrt{23/18}\) and \(\sigma=\sqrt2\), respectively. Dotted lines mark the total charge in each limiting core.}{fig:collapse}
\fig{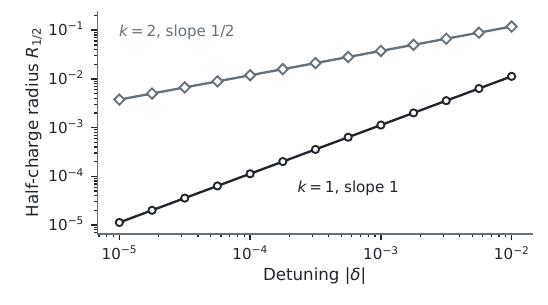}{0.82}{Half-charge radii for the aligned input at \(\tau=1\). Circles and diamonds give the radii solving \(Q_k(R,\delta)=k/2\) for one and two stages using the exact transfer. Lines show the asymptotic laws in Eq.~\eqref{eq:radii}.}{fig:radii}

\section{Aperture weighting and synthetic counts}\label{app:counts}
\paragraph{Probability in a shrinking aperture.}
The success probability averaged over an input aperture of radius \(h\) is
\[
 \bar p_k(h)=\frac{\int_{|w|<h} I_Ap_k\,\dd^2w}{\int_{|w|<h} I_A\,\dd^2w}.
\]
For a smooth incident density with \(I_A(0)>0\), the leading radial and angular weight is uniform over the disc. With Eq.~\eqref{eq:local}, the leading cross term averages to zero in angle and the disc average of \(|w|^{2k}\) is \(h^{2k}/(k+1)\). Thus
\[
 \bar p_k(h)=s_0^{2k}\left[D|\delta|^2+\frac{A}{k+1}h^{2k}
 + o(|\delta|^2+h^{2k})\right].
\]
The correction to the cross term from the spatial weight is \(O(h^{k+1}|\delta|)\), smaller than the leading terms. Setting \(h\) proportional to \(|\delta|^{1/k}\) gives \(\bar p_k(h)\) of order \(|\delta|^2\). For example, at \(\tau=1\), \(s_0=0.2\), and \(|\delta|=10^{-4}\), a two-stage aperture set to the leading half-charge radius has \(\bar p_2\simeq4.27\times10^{-11}\). Even two million trials within that aperture would give about \(8.5\times10^{-5}\) expected signal events.

\paragraph{Simulation and charge inference.}
Figure~\ref{fig:counts} uses a separate exposure with \(2\times10^6\) trials per stage and setting, allocated equally across 25 positions on a \(5\times5\) grid in \([-0.18,0.18]^2\), including the endpoints. The independent Poisson count in bin \(j\) has mean \(\mu_j=N_j[p_k(w_j)+10^{-5}]\), where \(N_j\) is the number of trials at that position. The additive term represents calibrated background and gives 20 expected background counts in total. We fit real nonnegative detuning with the transfer and input alignment specified by the model.

At each candidate \(\delta_j\), the likelihood-ratio statistic is \(L_j=2[\ell(\widehat\delta)-\ell(\delta_j)]\), with the Poisson log likelihood maximized over the same discrete grid for observed and simulated data. The fitting grid contains zero, 240 logarithmically spaced points from \(10^{-5}\) to \(0.2\), and the four generating values \(0.002,0.008,0.025,0.10\). We simulate 999 datasets at each candidate. The Monte Carlo \(p\)-value is \((1+n_{\ge})/1000\), where \(n_{\ge}\) is the number of simulated statistics at least as large as the observed statistic. We keep candidates with \(p>0.05\). This construction controls the rejection probability at each grid point under the count model. The interval plotted in Fig.~\ref{fig:counts} runs from the smallest to the largest disc charge among the retained candidates. It can span gaps in the set of projected charges because the charge map jumps at zero detuning.

For the aligned two-stage run at \(\delta=0.002\), the mean is 24.33 counts and the exact disc charge is 1.91627. The simulated 26 counts give an interval spanning essentially \([0,2]\). At \(\delta=0.10\), the mean is 88.74 counts and the exact disc charge is 0.01042. The simulated 101 counts give a fitted charge of about 0.00975 with interval \([0.00878,0.01054]\).

\section{Algebraic geometry aspect of dark points}\label{app:sheaf}
\paragraph{Input and output ray bundles.}
The biphoton entangled state can be written as
\[
 |\Psi\rangle=\mathcal N\int \dd^2w\,\chi(w)\,
 |w\rangle_A\otimes
 \left(|e_1\rangle_B+w|e_2\rangle_B+w^2|e_3\rangle_B\right),
\]
where \(\mathcal N\) normalizes the state. Detecting photon A at position \(w\) prepares photon B in the ray represented by
\(f(w)=(1,w,w^2)^{\mathsf T}.\)
Thus \(w\mapsto[f(w)]\) describes a family of conditional quantum states of photon B parametrized by the detection position of photon A. On the compactified position plane \(X=\CP^1\), these rays define the holomorphic line subbundle
\(L\cong\mathcal O(-2)\subset E=\mathcal O^{\oplus3},\)
where \(E\) is the trivial three-mode bundle of photon B. Here \(f(w)\) is a local frame on the finite chart; in the chart \(v=1/w\), the regular frame is \(v^2f(1/v)=(v^2,v,1)^{\mathsf T}\). When photon B passes through the tuned EP3 filter
\[
K_k=T_{k,0}=s_0^k\left(e^{-\ii\tau J_0}-I\right)^k, \qquad k=1,2,\quad\tau\ne0,
\]
its transmitted amplitude has the common factor \(w^k\). Cancelling this factor and extending the output ray through the dark point gives
\(L'_k\cong\mathcal O(k-2)\subset E.\)
For each fixed stage \(k\), write \(K=K_k\) and \(L'=L'_k\).
Denote the inclusions by \(a,a'\) and the quotient maps by \(q,q'\). The ray bundles and the optical map fit into the diagram
\begin{equation}
\begin{array}{ccccccccc}
&&0\\
&&\big\downarrow\\
0 &\longrightarrow& L &\xrightarrow{a}& E &\xrightarrow{q}& N=E/L &\longrightarrow&0\\
&& \big\downarrow\rlap{$\scriptstyle K|_L$}
&& \big\downarrow\rlap{$\scriptstyle K$}
&& \big\downarrow\rlap{$\scriptstyle \overline K$}\\
0 &\longrightarrow& L' &\xrightarrow{a'}& E &\xrightarrow{q'}& N'=E/L' &\longrightarrow&0\\
&&\big\downarrow\rlap{$\scriptstyle \pi$}\\
&&\mathcal T\\
&&\big\downarrow\\
&&0 .
\end{array}
\label{eq:ray-diagram}
\end{equation}
The two horizontal rows in Eq.~\eqref{eq:ray-diagram} describe the input and extended output ray bundles of the conditional photon state.

At each position, the inclusion \(a:L\hookrightarrow E\) embeds the input ray line into photon B's three-dimensional mode space. Taking the fiberwise quotient by this line gives the rank-two bundle \(N=E/a(L)\), and hence the upper short exact sequence. The inclusion \(a':L'\hookrightarrow E\) and the quotient \(N'=E/a'(L')\) give the corresponding sequence for the extended output ray. These quotients are bundles over the position sphere \(X\), rather than quotients of \(\CP^2\) by the traced surface. Both horizontal rows are short exact sequences of holomorphic bundles. The induced quotient map \(\overline K([v])=[Kv]\) is well defined because \(K(L)\subseteq L'\). The quotient describes mode directions transverse to the occupied ray; the spatial variation of the ray into these directions determines its Berry curvature. The mode inner product gives these directions their orthogonal representatives.

To compute the curvature flux, use the mode inner product to identify \(N\) or \(N'\) with the orthogonal complement of the corresponding ray, as a smooth Hermitian bundle; this identification is generally not holomorphic. If \(u\) is a local unit vector spanning that ray, its projector is \(P=uu^\dagger\), and \(P_\perp=I-P\) projects onto the orthogonal complement. Projecting the trivial connection of \(E\) onto the ray gives \(\nabla=P\dd\), with curvature \(P\dd P\wedge\dd P\,P\). Equivalently, the Berry connection is \(\mathcal A=-\ii u^\dagger\dd u\). The curvature \(\dd\mathcal A\) is the pullback of the projective-state-space curvature to \(X\); its flux measures the charge of the conditional-state texture. With the sign convention in Appendix~\ref{app:geometry},
\begin{equation}
 Q=\frac{1}{2\pi\ii}\int_X
 \operatorname{tr}\!\left(P\,\dd P_\perp\wedge\dd P_\perp\,P\right).
 \label{eq:quotient-charge}
\end{equation}
Indeed, \(\dd P_\perp=-\dd P\). For a local unit ray vector \(u\), the trace in Eq.~\eqref{eq:quotient-charge} is \(\dd(u^\dagger\dd u)=\ii\dd\mathcal A\). The first Chern number of the ray bundle is the negative of the charge in this convention \cite{r19}. Additivity of the first Chern class in each row, with \(E\) trivial, gives \(Q_{\rm in}=-\deg L=\deg N\) and \(Q_{\rm ext}=-\deg L'=\deg N'\). These degree identities give the integrated charges. Their local curvature densities follow from the spatial dependence of the projectors.

\paragraph{Quotient length and charge deficit.}
At the dark point, the transmitted amplitude vanishes even though the output ray has a well-defined limit. To separate these two facts, take local nonvanishing frames \(s,\tilde s\) and write \(Ka(s)=g\,a'(\tilde s)\). At the aligned dark point we can choose \(g=w^k\): the scalar factor gives the order of the amplitude zero, while \(a'(\tilde s)\) specifies the limiting ray. Both ray bundles have rank one. The full map \(K:E\to E\) has constant rank \(3-k\), but its restriction \(K|_L:L\to L'\) drops from rank one to zero at \(w=0\), where \(L_0\subseteq\ker K\). Thus the restriction remains a holomorphic bundle morphism, while its fiberwise cokernel cannot form a vector bundle of constant rank.

Passing to sheaves allows the cokernel at the dark point to be included in an exact sequence while retaining the order of vanishing of the transmitted amplitude. The completed left column in Eq.~\eqref{eq:ray-diagram} is therefore interpreted as a sequence of sheaves, with \(\pi:L'\to\mathcal T=L'/K(L)\) the cokernel map. Changes of local frame multiply \(g\) by a nowhere vanishing holomorphic function, so its zeros and their orders define an effective divisor \(D_0\) on \(X\). Here \(D_0=kp\), with \(p\) the point \(w=0\). The image sheaf is \(L'(-D_0)\). Multiplication by \(g\) is injective on holomorphic germs even though the fiber map vanishes at \(p\); hence the sequence is exact as sheaves, not fiberwise as vector bundles. With \(L,L'\) now denoting their sheaves of local holomorphic sections,
\begin{equation}
 0\longrightarrow L\xrightarrow{K}L'\longrightarrow \mathcal T\longrightarrow0,
 \qquad \mathcal T=L'\otimes\mathcal O_{D_0}.
 \label{eq:short-seq}
\end{equation}
The algebraic quantity needed to count the dark-point multiplicity is the length of this torsion cokernel. In the output frame, the local quotient is \(\mathbb C\{w\}/(w^k)\), where \(\mathbb C\{w\}\) is the ring of convergent power series at the origin. Multiplication by \(w^k\) cannot produce the Taylor coefficients of \(1,w,\ldots,w^{k-1}\); these are precisely the coefficients retained by the quotient. The length of the finite-support sheaf \(\mathcal T\) is the sum of the module lengths of its stalks over their local rings \cite{r28}. Since the residue fields are \(\mathbb C\), each stalk's length equals its complex dimension. Thus the first and second stages give lengths one and two, distinguishing the orders of the dark points even though both fiber maps drop rank by one.

This local multiplicity also determines the global charge deficit. The vanishing factor relates the input and extended output line bundles by \(L'\otimes L^*\cong\mathcal O(D_0)\) \cite{r28}. Combining this relation with the degree--charge identities from the horizontal sequences gives
\begin{equation}
 Q_{\rm in}-Q_{\rm ext}
 =\deg L'-\deg L
 =\deg D_0
 =\operatorname{length}(\mathcal T)=k .
 \label{eq:length-charge}
\end{equation}
This is the algebraic form of the winding calculation in Eq.~\eqref{eq:index}. It expresses the deficit in an integrated geometric quantity in terms of the local order of vanishing.

For sufficiently small nonzero detuning, define \(L'_\delta=T_{k,\delta}(L)\). The transfer restricts to an isomorphism \(L\to L'_\delta\), so its cokernel vanishes. Here \(L'_\delta\cong\mathcal O(-2)\) is the detuned output line, not the tuned extended line \(L'\cong\mathcal O(k-2)\). The third tuned stage extinguishes the field and supplies no output ray line.

\FloatBarrier

\end{document}